\documentclass[runningheads]{llncs}

\usepackage{eccv}

\usepackage{eccvabbrv}

\usepackage{graphicx}
\usepackage{booktabs}

\usepackage{tikz}

\usepackage[accsupp]{axessibility}  

\usepackage{hyperref}

\usetikzlibrary{3d,decorations.text,shapes.arrows,positioning,fit,backgrounds,calc,shapes.geometric}

\tikzset{
    pics/fake box/.style args={#1 with dimensions #2 and #3 and #4}{
        code={
            \draw[gray,ultra thin,fill=#1]  (0,0,0) coordinate(-front-bottom-left)
            to ++ (0,#3,0) coordinate(-front-top-left)
            --++ (#2,0,0) coordinate(-front-top-right)
            --++ (0,-#3,0) coordinate(-front-bottom-right) -- cycle;
            \draw[gray,ultra thin,fill=#1] (0,#3,0)  --++ (0,0,#4) coordinate(-back-top-left)
            --++ (#2,0,0) coordinate(-back-top-right) --++ (0,0,-#4)  -- cycle;
            \draw[gray,ultra thin,fill=#1!80!black] (#2,0,0) --++ (0,0,#4) coordinate(-back-bottom-right)
            --++ (0,#3,0) --++ (0,0,-#4) -- cycle;
            \path[gray,decorate,decoration={text effects along path,text={CONV}}]
            (#2/2,{2+(#3-2)/2},0) -- (#2/2,0,0);
        }
    }
}

\tikzset{circle dotted/.style={dash pattern=on .05mm off 2mm, line cap=round}}

\usepackage{orcidlink}

\usepackage{multirow}

\begin{document}

\title{Better Spanish Emotion Recognition In-the-wild: Bringing Attention to Deep Spectrum Voice Analysis}

\author{Elena Ortega-Beltrán\inst{1}\and
Josep Cabacas-Maso \inst{1}\and
Ismael Benito-Altamirano\inst{1,2}\orcidlink{0000-0002-2504-6123} \and Carles Ventura \inst{1}}


\institute{eHealth Center, Faculty of Computer Science, Multimedia and Telecommunicactions, Universitat Oberta de Catalunya, 08016 Barcelona, Spain
\email{ibenitoal@uoc.edu}, \email{cventuraroy@uoc.edu} \\ \and
MIND/IN2UB, Department of Electronic and Biomedical Engineeering, Universitat de Barcelona, 08028 Barcelona, Spain \\
}

\authorrunning{Elena Ortega-Beltrán et al.}
\titlerunning{Better Spanish Emotion Recognition In-the-wild}

\maketitle

\begin{abstract}

    Within the context of creating new Socially Assistive Robots, emotion recognition has become a key development factor, as it allows the robot to adapt to the user's emotional state in the wild. In this work, we focused on the analysis of two voice recording Spanish datasets: ELRA-S0329 and EmoMatchSpanishDB. Specifically, we centered our work in the paralanguage, e.~g. the vocal characteristics that go along with the message and clarifies the meaning. We proposed the use of the DeepSpectrum method, which consists of extracting a visual representation of the audio tracks and feeding them to a pretrained CNN model. For the classification task, DeepSpectrum is often paired with a Support Vector Classifier --DS-SVC--, or a Fully-Connected deep-learning classifier --DS-FC--. We compared the results of the DS-SVC and DS-FC architectures with the state-of-the-art (SOTA) for ELRA-S0329 and EmoMatchSpanishDB. Moreover, we proposed our own classifier based upon Attention Mechanisms, namely DS-AM. We trained all models against both datasets, and we found that our DS-AM model outperforms the SOTA models for the datasets and the SOTA DeepSpectrum architectures. Finally, we trained our DS-AM model in one dataset and tested it in the other, to simulate real-world conditions on how biased is the model to the dataset.
\end{abstract}

\section{Introduction}

During the last decades, an increasing of age population has been observed in many countries, i.~e. the USA~\cite{preston2021changing} or Western Europe countries~\cite{vsidlo2020retrospective}, this has let to the proposal for Socially Assistive Robots (SAR) to help elder people with day-to-day problems, but also aiming to help in with mental struggles related to age~\cite{yu2022socially}. Within this context, human emotion recognition becomes a key factor in the successful adoption of such technologies, and proposals have came around using artificial intelligence to solve this problem ~\cite{abdollahi2022artificial}.

In this work, we focused on the analysis of voice-recorded data from Spanish speakers, as Spanish is one of the most spoken languages in Europe, South and North America and often does not receive the same attention from researchers than English. We selected two databases: (1) ELRA-S0329~\cite{kerkeni_speech_2018}, a stock dataset from the European Language Resources Association which contains recordings of professional speakers in six emotions (anger, disgust, fear, joy, sadness, surprise) plus a neutral style in fast, slow, soft, loud and normal style, and (2) EmoMatchSpanishDB~\cite{garcia-cuesta_emomatchspanishdb_2023}, a dataset from the \emph{Universidad Europea de Madrid} which contains recordings of 50 individuals expressing six different emotions (anger, disgust, fear, happiness, sadness, surprise) and a neutral one. 

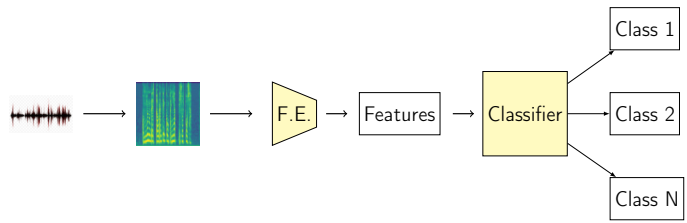
\begin{figure}[h!]
    \centering
    \resizebox{0.75\textwidth}{!}{
    \begin{tikzpicture}[x={(1,0)}, y={(0,1)}, z={({cos(60)},{sin(60)})},
    font=\sffamily\Large,scale=1]

        \node (audio) at (0,0) {\includegraphics[width=1.5cm]{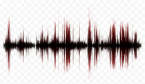}};
        \node at (0, -1) {};

        \draw[->, thick] (1,0) -- (2,0);

        \node (mel) at (3,0) {\includegraphics[width=1.5cm]{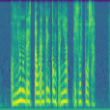}};
        \node at (3, -1) {};

        \draw[->, thick] (4,0) -- (5,0);

        \node[draw, trapezium, trapezium stretches=true, fill=yellow!30, minimum width=1.5cm, minimum height=1cm, trapezium angle=70, shape border rotate=270] (mfn) at (6,0) {F.E.};

        \draw[->, thick] (6.75,0) -- (7.25,0);

        \node[draw, fill=white, minimum width=1.5cm, minimum height=1cm, right=1cm of mfn] (features) {Features};

        \draw[->, thick] (9.75,0) -- (10.25,0);

        \node[draw, fill=yellow!30, minimum width=2cm, minimum height=2cm, right=1cm of features] (classifier) {Classifier};

        \node[draw, fill=white, minimum width=1.5cm, minimum height=1cm, above right=0.5cm and 1cm of classifier] (class1) {Class 1};
        \node[draw, fill=white, minimum width=1.5cm, minimum height=1cm, right=1cm of classifier] (class2) {Class 2};
        \node[draw, fill=white, minimum width=1.5cm, minimum height=1cm, below right=0.5cm and 1cm of classifier] (classN) {Class N};

        \draw[-latex] (classifier) -- (class1);
        \draw[-latex] (classifier) -- (class2);
        \draw[-latex] (classifier) -- (classN);

        \end{tikzpicture}
    }
    \caption{Schematic of a DeepSpectrum pipeline for audio classification. The audio signal is converted into a Mel spectrogram, which is then fed to a pretrained CNN backbone to extract features. The features are then classified using an available classifier.}
    \label{fig:audio_classification}
\end{figure}

We proposed to follow the work of Amiriparian et al., using DeepSpectrum toolkit~\cite{amiriparian_snore_2017} to extract paralinguistic features from the audio data using a pipeline that consisted of: (1) converting audio data into a spectrum representation of this data --a Mel spectrogram--; (2) using pretrained a VGG16 CNN~\cite{simonyan2014very} as an image feature extractor; and, (3) using a Support Vector Classifier (SVC) to classify the emotions for each of the datasets. We named this architecture DeepSpectrum-SVC  or, simply, DS-SVC (see~\autoref{fig:audio_classification}). We trained this architecture under a 10-fold cross-validation process and compared the results with the state-of-the-art models for each dataset~\cite{kerkeni_speech_2018,garcia-cuesta_emomatchspanishdb_2023}.

Additionally, we proposed to use a well-known technique to solve the classification task, using a Fully-Connected deep-learning classifier, which we named DeepSpectrum-FC or, simply, DS-FC. Moreover, we proposed our own classifier based on Attention Mechanisms, following the work of Gorriz et al.~\cite{gorriz2019assessing}. We named this architecture DeepSpectrum-AM or, simply, DS-AM. We trained all models against both datasets and we found that our DS-AM model outperforms the SOTA models for the datasets and the SOTA DeepSpectrum architectures.

Finally, we trained our DS-AM model in one dataset and tested it in the other, in order to simulate in-the-wild conditions where the audio samples are not only from unknown speakers but also with other acoustic conditions and different input texts.

All in all, the DS-SVC architecture performed better than its counterparts for both datasets; the DS-FC only outperformed SOTA models that used an SVC to solve the classification task; and, the DS-AM model outperformed all SOTA models for both  datasets and the DS-SVC and DS-FC architectures that we also studied. The DS-AM model was trained in one dataset and tested in the other showed that the model is biased to the dataset, as it performed worse than the DS-AM model trained and tested in the same dataset. And as expected, the EmoMatchSpanishDB dataset, having 50 speakers instead of the 2 in ELRA-S0329, is more appropriate for in-the-wild emotion recognition.

\section{Related Work}

\subsection{Speech databases}

Emotion recognition is a widely studied field that comprises several disciplines, such as psychology or linguistics, and in recent decades, also computer science. Normally, speech databases introduce labels to classify or quantify the emotions expressed in the recordings. These labels can be discrete, such as the ones proposed by Ekman~\cite{ekman2004emotional}, which consists on several emotions, i. e. ``anger'' or ``joy''; or continuous, such as the valance-arousal system~\cite{Bestelmeyer2017}, which consists on two dimensions that describe the emotional state of the speaker.

In addition to this, speech databases distinguish themselves by the way the recordings are obtained. We can divide them in three categories:

\begin{itemize}
    \item \textbf{Acted or Simulated:} These databases are recorded by professional actors in a recording studio. Although they usually have very good audio quality, they are not so realistic or useful as real ones. Examples are: EmoDB~\cite{burkhardt2005database}, IEMOCAP~\cite{busso2008iemocap} Spanish Expressive Voices Corpus Description~\cite{barra2008spanish}, ELRA-S0329~\cite{kerkeni_speech_2018} or EmoMatchSpanishDB~\cite{garcia-cuesta_emomatchspanishdb_2023}.

    \item\textbf{ Induced:} There are created placing the actors in simulated situations to produce the requested emotion. They are more useful than the acted ones, as they are more close to real situations. An example is eNTERFACE’05~\cite{Martin2006}.

    \item \textbf{ Natural:} They are obtained from real situations as talk-shows, social media or call centers. Although the more useful of all, they have legal and ethical issues to use them.  Examples are RECOLA Speech Database.~\cite{ringeval2013introducing} or CMU-MOSEAS~\cite{cmumoseas2020}.
\end{itemize}

 

\subsection{Acoustic characteristics}

From a phonological point-of-view, we can distinguish between segmental and suprasegmental characteristics in the study of audio tracks. On one hand, segmental characteristics are tied to a limited window of time within an utterance (such as a phoneme, approximately 20-30ms). The most common characteristics are those associated with the \emph{cepstrum}, which consists of a transformation of the spectrum that allows us to discover periodic features in the frequency domain. For example, the MFCC (Mel Frequency Cepstral Coefficients) or the LPCC (Linear Prediction Cepstral Coefficients) belong to this category, and their definition can be found in the common literature~\cite{vary2023digital}.

On the other hand, the suprasegmentals characteristics are tied to the entire utterance as a whole, like speech rate, shimmer or jitter. Often, these characteristics are evaluated in much larger windows of time, these characteristics are called Low-Level Descriptors (LLD), a widely spread standard for these features is the ComParE~\cite{schuller2023acm} feature set, composed of 6373 static features derived from low-level descriptors (LLD). Also, another widespread feature set is the MSFs (Modulation Spectral Features~\cite{wu_automatic_2011}), which are processed through a set of filters that simulate the human auditory system.

The usual approach from authors to machine learning solutions is: first, to implement any sort of these above-mentioned classical feature extraction method and, then, use a classifier to solve the classification task. The classifier can be as simple as a Support Vector Classifier (SVC), or more complex, like a Recurrent Neural Network (RNN).

For the ELRA-S0329 dataset, Kerkeni et al.~\cite{kerkeni_speech_2018} used two different feature extractors: MSFs and MFCC; and they combined both, obtaining three different feature sets (MSF, MFCC and MSF+MFCC). Regarding the classifier task, they introduced three different classifiers, first a Multi-Linear Regression (MLR), then a Support Vector Machine classifier (SVC) and finally a Recurrent Neural Network (RNN).~\autoref{tab:elra_baseline} shows the results obtained by these authors.

\begin{table}[ht]
    \centering
    \caption{Accuracy results obtained for MLR, SVC, and RNN at ELRA-S0329 (MSF, MFCC and MSF+MFCC features sets)~\cite{kerkeni_speech_2018}.}
    \label{tab:elra_baseline}
    \begin{tabular}{lccc}
        \toprule
        & \textbf{MLR} & \textbf{SVC} & \textbf{RNN} \\

        \midrule
        MSF & 0.706 & 0.776 & 0.823 \\
        MFCC & 0.761 & 0.707 & 0.866 \\
        MSF+MFCC & 0.824 & 0.681 & 0.905 \\
        \bottomrule
    \end{tabular}
\end{table}

For the EmoMatchSpanishDB dataset, García-Cuesta et al.~\cite{garcia-cuesta_emomatchspanishdb_2023} used the well-known ComParE feature set for audio preprocessing, they also introduced the EgeMaps~\cite{lin2014emotion} feature set. Regarding the classifier task, they three algorithms: a Support Vector Machine Classifier (SVC), a XGBOOST classifier and a Feed-Foward Neural Network (FFNN).~\autoref{tab:emo_match_baseline} shows the results obtained by these authors.

\begin{table}[ht]
\centering
\caption{F1-Score and accuracy results obtained for SVC, XGBOOST, and FFNN at EmoMatchSpanishDB (eGeMAPS and CompaRE features sets)~\cite{garcia-cuesta_emomatchspanishdb_2023}.}
\label{tab:emo_match_baseline}
\begin{tabular}{l|cc|cc|cc}
    \toprule
     & \multicolumn{2}{c}{\textbf{SVC}} & \multicolumn{2}{c}{\textbf{XGBOOST}} & \multicolumn{2}{c}{\textbf{FFNN}} \\
    \cline{2-7}
    & \textbf{F1} & \textbf{Acc} & \textbf{F1} & \textbf{Acc} & \textbf{F1} & \textbf{Acc} \\
    \midrule
    eGeMAPS & 0.522 & 0.542 & 0.503 & 0.562 & 0.511 & 0.554 \\
    CompaRE & 0.589 & 0.642 & 0.588 & 0.642 & 0.567 & 0.596 \\
    \bottomrule
\end{tabular}
\end{table}

%
%


\subsection{DeepSpectrum}


DeepSpectrum arose as full deep-learning alternative to the classical feature extraction methods. Originally intended to extract features from snore human sounds~\cite{amiriparian_snore_2017}, currently, it presents itself as a Python toolkit that produces visual representation of audio tracks --spectrogram or chromagram-- and feeds them to a pretrained CNN for image feature extraction. Insted of extracting classical characteristics like the Mel Frequency Cepstral Coefficients (MFCC) or the Modulation Spectral Features (MSFs), DeepSpectrum uses the CNN to extract the features from the Mel spectrogram themselves.

It supports widely more than 10 different backbone models, such as VGG16, ResNet, DenseNet, Inception, etc. Plus, it allows to select from which convolutional layer the features will be extracted. Moreover, other parameters can be selected, like: window and hopsize for feature extraction, length of fft window used for creating the spectograms, frequency scales and limits, color map of the plots, etc.~\cite{deepspectrum2020}.

\begin{figure}[ht!]
    \centering
    \includegraphics[width=\textwidth,clip,trim=3em 0 3em 0]{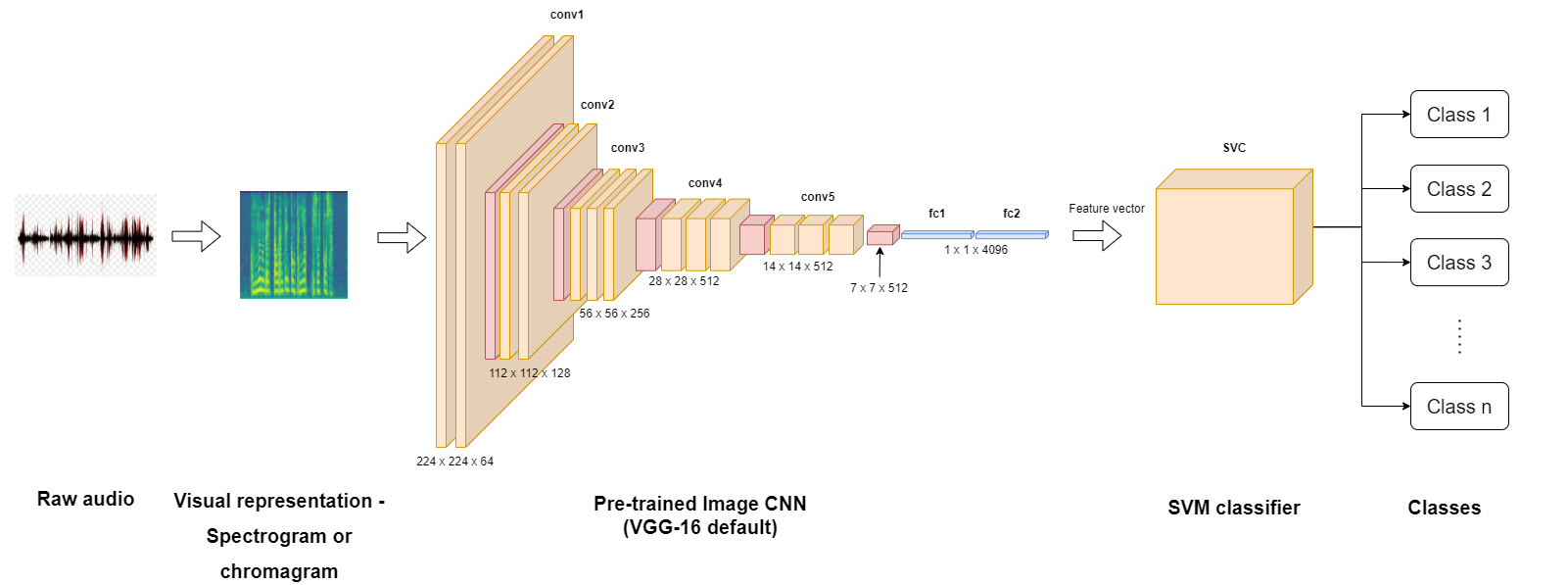}
    \caption{A typical DeepSpectrum-SVC (DS-SVC) pipeline for audio classification. The audio signal is converted into a Mel spectrogram, which is then fed to a pretrained CNN backbone to extract features. The features are then classified using a Support Vector Classifier.}
    \label{fig:deepspectrum1}
\end{figure}

In the original deep spectrum analysis, Amiriparian et al. used a Supported Vector Machine classifier to solve the classification task after extracting the features with the CNN, this is depicted at~\autoref{fig:deepspectrum1}.

\begin{figure}
    \centering
    \includegraphics[width=\textwidth,clip,trim=3em 0 3em 0]{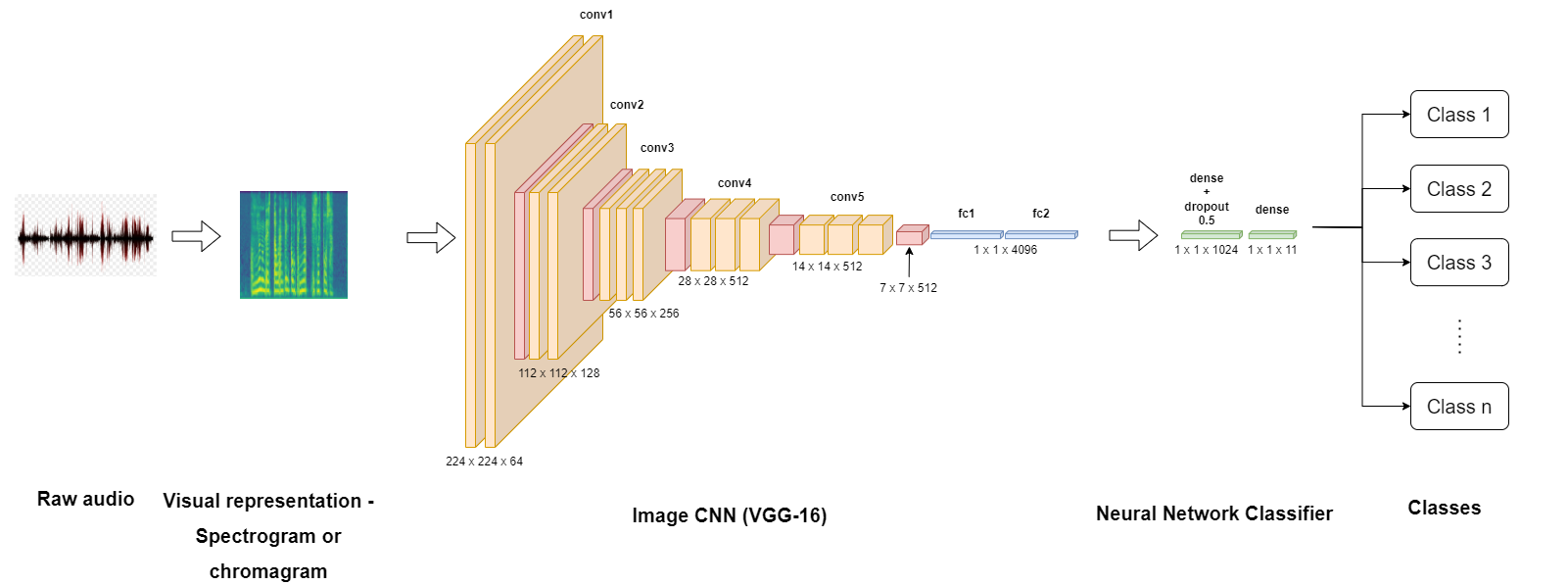}
    \caption{Another configuration of DeepSpectrum, DeepSpectrum-FC (DS-FN) where the features are extracted from the Mel spectrogram using a pretrained CNN backbone and then classified using a Fully-Connected deep-learning classifier.}
    \label{fig:deepspectrum2}
\end{figure}

In other related works, such as in JAafar and Lachiri ~\cite{jaafar2022stress}, were authors used a Fully-Connected deep-learning classifier to solve the classification task, this is depicted at~\autoref{fig:deepspectrum2}. Note this was also considered in the original DeepSpectrum work but only as a future work possibility~\cite{amiriparian_snore_2017}.

\section{Materials and Methods}

\subsection{Datasets}

Two datasets were considered in this study, both of them are in Spanish language:

\begin{itemize}
    \item \textbf{ELRA-S0329~\cite{kerkeni_speech_2018}} was presented on 2018 and consists on the recordings of one female and one male professional speakers in six emotions (anger, disgust, fear, joy, sadness, surprise) plus a neutral style in fast, slow, soft, loud and normal style. It contains 6041 archives with a total recording time of 7 hours and 52 minutes. In the original work, authors trained the dataset using a 10-cross validation scheme, first they splitted up the folds and later, they created a 70\% split for training and 30\% for testing. In order to compare to the state-of-the-art, we will use the same methodology, for this dataset. The best baseline obtained by Kerkeni et al. was a RNN with MFCC and MS features, with an accuracy of 90.05\%~\cite{kerkeni_speech_2018}, as is described in~\autoref{tab:elra_baseline}.

    \item \textbf{EmoMatchSpanishDB~\cite{garcia-cuesta_emomatchspanishdb_2023}} was presented on 2023 and contains the recordings of 50 individuals (31 male and 19 female) with a total of 2050 archives expressing six different emotions (anger, disgust, fear, happiness, sadness, surprise) and a neutral one. In the original work, authors introduced a cross validation scheme named ``Leave-One-Speaker-Out'' (LOSO), where they used 45 speakers for training and 5 speakers for testing. The best baseline obtained by García-Cuesta et al. was a SVC (or XGBOOST) classifier with the ComParE feature set, with an accuracy of 64.2\%~\cite{garcia-cuesta_emomatchspanishdb_2023}, as is described in~\autoref{tab:emo_match_baseline}.

\end{itemize}

There exists a noticeable difference in the baseline accuracy despite the similarities in the feature extraction pipelines. This is somehow to be expected, as the EmoMatchSpanishDB dataset has 50 speakers instead of the 2 in ELRA-S0329. Nevertheless, we understood this distinction as an opportunity to test the robustness of our models in different conditions.

\subsection{Proposed method}

\begin{figure}[ht!]
    \centering
    \begin{subfigure}[b]{\textwidth}
        \centering
        \caption*{a)}
        \includegraphics[width=\textwidth]{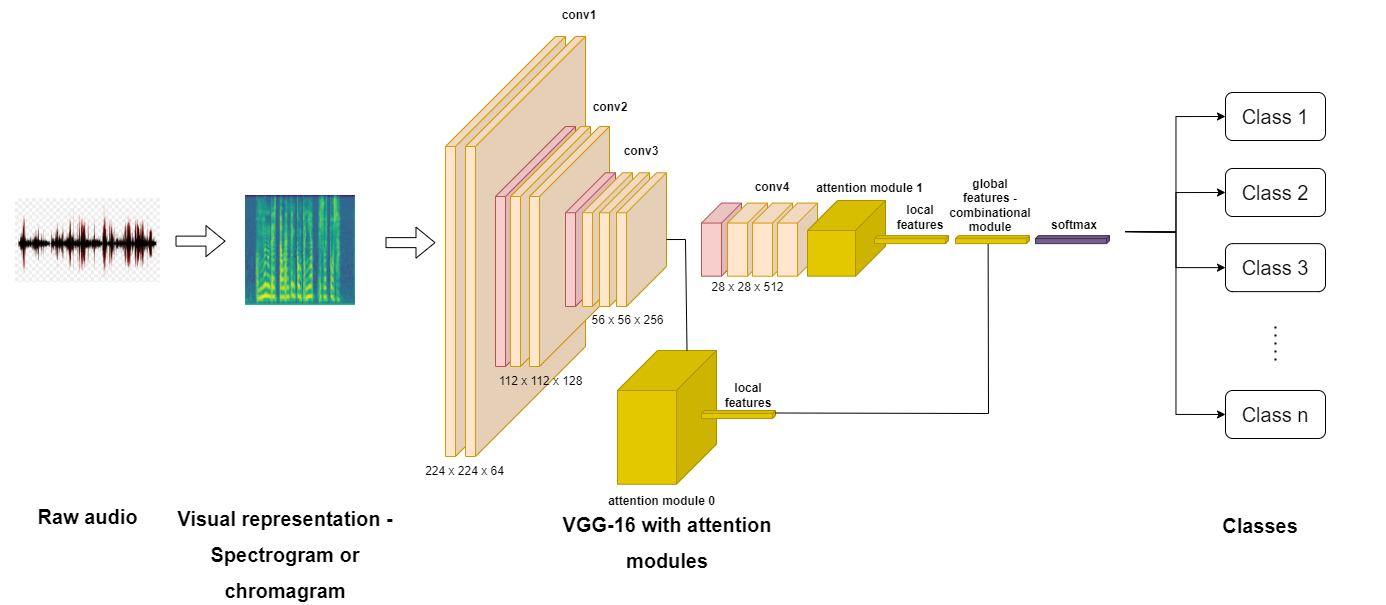}
    \end{subfigure}
    \vspace{1em}
    \begin{subfigure}[b]{0.65\textwidth}
        \centering
        \caption*{b)}
        \includegraphics[width=\textwidth]{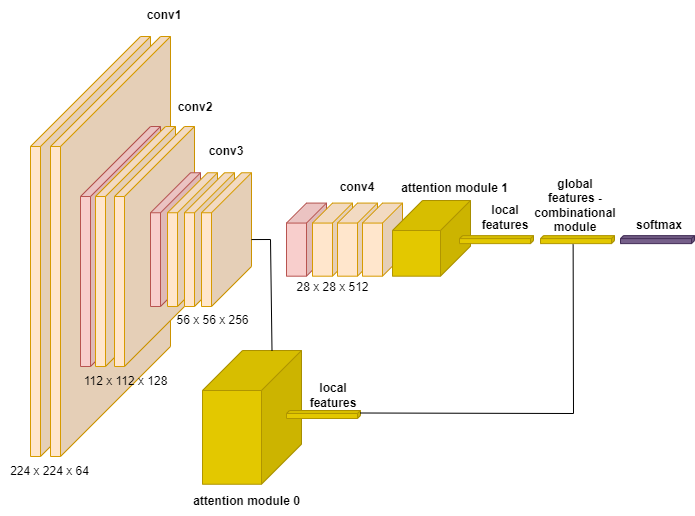}
    \end{subfigure}
    \caption{The DeepSpectrum-Attention Mechanisms (DS-AM) architecture. The figure shows: a) the general model which includes a modified DeepSpectrum-AM; and, b) the attention mechanism module.}
    \label{fig:vggatt1}
\end{figure}

We proposed to use our own variant of a DeepSpectrum pipeline, this approximation consists on bringing attention mechanisms to the DeepSpectrum architecture. Instead of using the entire back-bone CNN to extract the image features from the Mel spectrogram, we modified a VGG-16 CNN to include two attention mechanisms, following the work of Gorriz et al.~\cite{gorriz2019assessing}, who used this architecture to evaluate the severity of knee osteoarthritis processing X-Ray images. We named this architecture DeepSpectrum-AM or, simply, DS-AM. The attention mechanisms were added to the last two convolutional blocks of the VGG-16, as shown in~\autoref{fig:vggatt1}.

\subsection{Experimental design}

As the first experiment, we proposed a transfer learning solution to the classification task by using the pretrained VGG-16 CNN in the DeepSpectrum toolkit (from Keras stock library --where VGG-16 was trained over ImageNet~\cite{deepspectrum2020}--) to extract the features from the Mel spectrograms. Then we used the well-known SVM classifier, as proposed in the original work of DeepSpectrum~\cite{amiriparian_snore_2017}. We named the architecture for this experiment DeepSpectrum-SVC or, simply, DS-SVC (see~\autoref{fig:deepspectrum1}). Note that we used a VGG-16 CNN as the backbone, as it is the same architecture used in the further experiments, specially in the case of the addition of attention mechanisms.


As the second experiment, we proposed a fine-tuning solution for the feature extraction step paired with a fully-connected deep-learning. For the fine-tuning, we obtained the same pretrained stock network --VGG-16 trained on ImageNet from Keras-- and we unfroze all layers with a low-learning rate. We expected this approximation to better handle the specificity of our datasets, as our images are Mel spectograms, not general images. We named this architecture DeepSpectrum-FC or, simply, DS-FC (see~\autoref{fig:deepspectrum2}).

As third experiment, we proposed a step further in network configuration, by taking a modified version of the VGG-16 network with two attention mechanisms, following the work of Gorriz et al.~\cite{gorriz2019assessing}. As we took the VGG-16 from the one pretrained in Keras, we used all the pretrained weights for the firsts layers of the extractor, but start using new untrained values for the attention heads. We named this architecture DeepSpectrum-Attention Mechanisms or, simply, DS-AM.

%
%

For experiments 1, 2 and 3, for ELRA-S0329, we have grouped all the five neutral labelled emotions in only one category, as we consider that the neutral emotion contained in EmoMatchSpanishDB only represents the normal category and doesn't contain the wide range that ELRA-S0329 has. We trained over both datasets, ELRA-S0329~\cite{kerkeni_speech_2018} and EmoMatchSpanishDB~\cite{garcia-cuesta_emomatchspanishdb_2023}, following the partitions that most resembled the original works. This means that for both datasets we used a 10-fold cross-validation process, as described in the original works. Despite this, notice that this means that the folds for the ELRA-S0329 dataset are constituted by different instances from the two same speakers, and the folds for the EmoMatchSpanishDB dataset are constituted by different instances from the 50 speakers. In this case, as we have a total of 50 speakers, we used 45 speakers for training and 5 speakers for validation.

As the fourth experiment, we trained the best model obtained with each of the datasets and we tested its performance in the other dataset, we did not use the cross-validation here. Doing so, we simulated an in-the-wild scenario where our task should be applied, being the audio samples not only from unknown speakers but also with other acoustic conditions and different input texts. For this experiment, for ELRA-S0329, we used only the normal-neutral labeled samples for testing purposes, although we maintained the whole set of five neutral emotions for training. We considered that the neutral emotion contained in EmoMatchSpanishDB only represents the normal category and doesn't contain the wide range that ELRA-S0329 has.

%

\section{Experimental Results}

\autoref{tab:experiment_1_results} shows the results of training DeepSpectrum-SVC over both datasets, ELRA-S0329 and EmoMatchSpanishDB, and compares them with the state-of-the-art models for each dataset regarding machine-learning classic classifiers. It can be shown that our model performed similarly to the SOTA models for the ELRA-S0329 dataset, but it was outperformed by the SOTA models for the EmoMatchSpanishDB dataset.

\begin{table}[ht!]
    \centering
    \caption{Comparison between the SOTA models for machine-learning classification solvers: MRL and SVC for ELRA-S0329 dataset~\cite{kerkeni_speech_2018}; and SVC and XGBOOST for EmoMatchSpanishDB dataset~\cite{garcia-cuesta_emomatchspanishdb_2023}. Our model using a pretrained VGG-16 CNN as feature extractor and a SVC as classifier is also shown, which scores similar to SOTA for the ELRA-S0329 dataset.}
    \label{tab:experiment_1_results}
    \begin{tabular}{lcccc}
        \toprule
        \textbf{Feature Set} & \textbf{Classifier} & \textbf{Acc.$_{\mathrm{ELRA-S0329}}$} & \textbf{Acc.$_{\mathrm{EmoMatchSpanishDB}}$} \\
        \midrule
        \multirow{2}{*}{MSF}      & MLR & 0.706 & - \\
                                  & SVC & 0.776 & - \\
        \multirow{2}{*}{MFCC}     & MLR & 0.761 & - \\
                                  & SVC & 0.707 & - \\
        \multirow{2}{*}{MSF+MFCC} & MLR & \textbf{0.824} & - \\
                                  & SVC & 0.681 & - \\
        \midrule
        \multirow{2}{*}{EgeMaps}  & SVC & - & 0.542 \\
                                  & XGBOOST & - & 0.562 \\
        \multirow{2}{*}{ComParE}  & SVC & - & \textbf{0.642} \\
                                  & XGBOOST & - & \textbf{0.642} \\
        \midrule
        \multicolumn{2}{c}{DeepSpectrum-SVC (DS-SVC)} & \textbf{0.821} & 0.525 \\
        \bottomrule
    \end{tabular}
\end{table}

\autoref{tab:experiment_2_3_results} show the results of training DeepSpectrum-FC and DeepSpectrum-AM over both datasets, ELRA-S0329 and EmoMatchSpanishDB --experiments 2 and 3--, and compares them with the state-of-the-art models for each dataset regarding deep-learning classifiers. It can be shown that our model outperformed the SOTA models for both datasets and the SOTA DeepSpectrum architectures. Specially for the EmoMatchSpanishDB dataset, can be seen that our deep-learning models surpassed the XGBOOST classifier, which was the best model in the original work, see also~\autoref{tab:experiment_1_results}.

\begin{table}[ht!]
    \centering
    \caption{Comparison between the SOTA models for deep-leaning classification solvers: RNN for ELRA-S0329 dataset~\cite{kerkeni_speech_2018}; FFNN for EmoMatchSpanishDB dataset~\cite{garcia-cuesta_emomatchspanishdb_2023}; and our model using a pretrained VGG-16 CNN as feature extractor and: a Fully-Connected deep-learning classifier (DS-FC) and a CNN with two attention mechanisms (DS-AM).}
    \label{tab:experiment_2_3_results}
    \begin{tabular}{lcccc}
        \toprule
        \textbf{Feature Set} & \textbf{Classifier} & \textbf{Acc.$_{\mathrm{ELRA-S0329}}$} & \textbf{Acc.$_{\mathrm{EmoMatchSpanishDB}}$} \\
        \midrule
        MSF                       & RNN & 0.823 & - \\
        MFCC                      & RNN & 0.866 & - \\
        MSF+MFCC                  & RNN & 0.905 & - \\
        \midrule
        EgeMaps                   & FFNN & - & 0.554 \\
        ComParE                   & FFNN & - & 0.554 \\
        \midrule
        \multicolumn{2}{c}{DeepSpectrum-FC (DS-FC)} & \textbf{0.975} & \textbf{0.659} \\
        \multicolumn{2}{c}{DeepSpectrum-AM (DS-AM)} & \textbf{0.984} & \textbf{0.683} \\
        \bottomrule
    \end{tabular}
\end{table}

\begin{figure}[ht!]
\centering
    \begin{subfigure}[b]{0.48\textwidth}
        \centering
        \caption*{a)}
        \includegraphics[width=\textwidth]{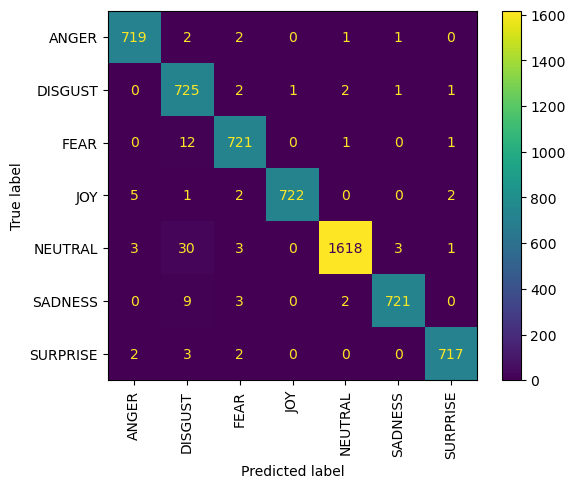}
    \end{subfigure}
    \vspace{1em}
    \begin{subfigure}[b]{0.48\textwidth}
        \centering
        \caption*{b)}
        \includegraphics[width=\textwidth]{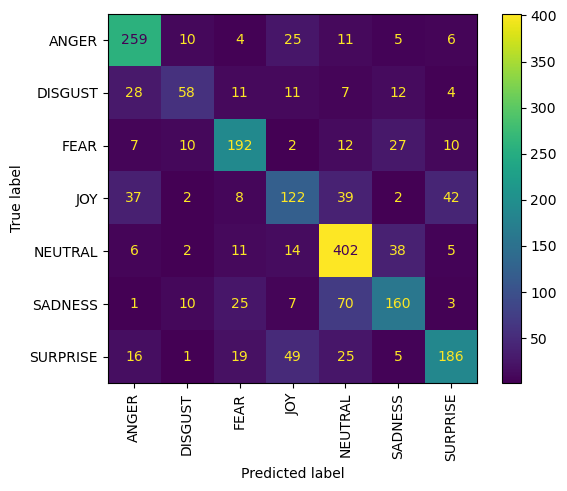}
    \end{subfigure}
    \caption{Confusion matrix for DeepSpectrum-AM over a) ELRA-S0329 dataset; and, b) EmoMatchSpanishDB dataset.}
\label{fig:cm}
\end{figure}

\newpage
Additional results are shown in~\autoref{tab:table2} where a detailed classification report is shown for the ELRA-S0329 dataset and DeepSpectrum-AM model.~\autoref{tab:table4} shows the same for the EmoMatchSpanishDB dataset and DeepSpectrum-AM model.  For this architecture, also a detailed view of the confusion matrix for both datasets is shown in~\autoref{fig:cm}.

\begin{table}[!ht]
    \centering
    \caption{Classification report for DeepSpectrum-AM model trained over the ELRA-S0329 dataset.}
    \label{tab:table2}
    \begin{tabular}{lccc}
        \toprule
        \textbf{Label} & \textbf{Precision} & \textbf{Recall} & \textbf{F1-score} \\
        \midrule
        ANGER    & 0.986   & 0.992   & 0.989 \\
        DISGUST  & 0.927   & 0.991   & 0.958 \\
        FEAR     & 0.980   & 0.978   & 0.979 \\
        JOY      & 0.996   & 0.985   & 0.990 \\
        NEUTRAL  & 0.996   & 0.977   & 0.986 \\
        SADNESS  & 0.994   & 0.981   & 0.987 \\
        SURPRISE & 0.994   & 0.989   & 0.991 \\
        \bottomrule
    \end{tabular}
\end{table}

\begin{table}[!ht]
    \centering
    \caption{Classification report for DeepSpectrum-AM model trained over the EmoMatchSpanishDB dataset.}
    \label{tab:table4}
    \begin{tabular}{lccc}
        \toprule
        \textbf{Label} & \textbf{Precision} & \textbf{Recall} & \textbf{F1-score} \\
        \midrule
        ANGER    & 0.732   & 0.809   & 0.769 \\
        DISGUST  & 0.624   & 0.443   & 0.518 \\
        FEAR     & 0.711   & 0.738   & 0.724 \\
        JOY      & 0.530   & 0.488   & 0.508 \\
        NEUTRAL  & 0.710   & 0.841   & 0.770 \\
        SADNESS  & 0.643   & 0.580   & 0.610 \\
        SURPRISE & 0.727   & 0.618   & 0.668 \\
        \bottomrule
    \end{tabular}
\end{table}



\begin{figure}[ht!]
\centering
    \begin{subfigure}[b]{0.48\textwidth}
        \centering
        \caption*{a)}
        \includegraphics[width=\textwidth]{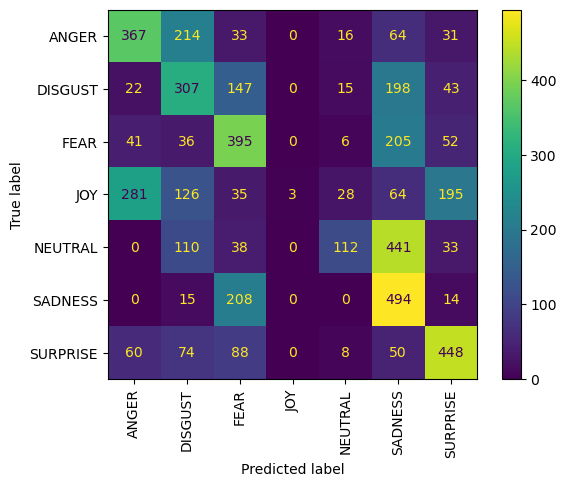}
    \end{subfigure}
    \vspace{1em}
    \begin{subfigure}[b]{0.48\textwidth}
        \centering
        \caption*{b)}
        \includegraphics[width=\textwidth]{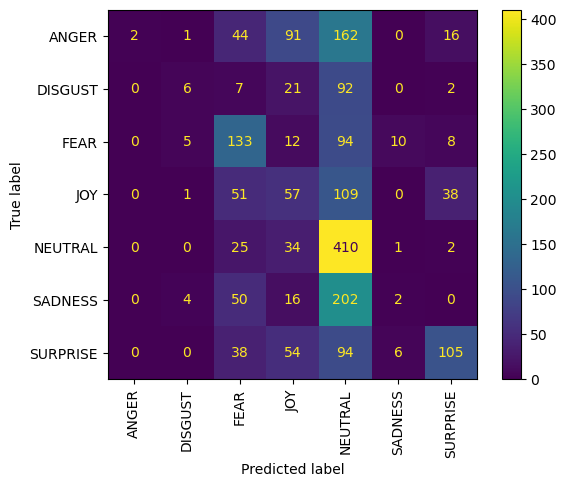}
    \end{subfigure}
    \caption{Confusion matrix for DeepSpectrum-AM model: a) training in EmoMatchSpanishDB and testing in ELRA-S0329; and, b) training in ELRA-S0329 and testing in EmoMatchSpanishDB.}
\label{fig:cm2}
\end{figure}

Results for experiment 4 are shown in~\autoref{tab:table4b}. It can be seen that the model trained in ELRA-S0329 and tested in EmoMatchSpanishDB performed poorly, while the model trained in EmoMatchSpanishDB and tested in ELRA-S0329 performed better. This is also shown in the confusion matrices in~\autoref{fig:cm2}. This is considered a good result, and to be expected. As training in EmoMatchSpanishDB, the model has to learn to generalize the features of the emotions, and testing on in-the-wild scenario --ELRA-S0329-- the model has to learn to generalize the features of the speakers.

\newpage
Additional results for the in-the-wild experiment can be found in the~\autoref{tab:table6} and~\autoref{tab:table7}, where the confusion matrices for both crossed experiments can be found in~\autoref{fig:cm2}. Finally, a detailed report for precision, recall and F1-score for both experiments can be found in~\autoref{tab:table6} and~\autoref{tab:table7}.


\begin{table}[!t]
    \centering
    \caption{Accuracy comparison between training and testing in ELRA-S0329 and EmoMatchSpanishDB with VGG16 + 2 attention mechanisms}
    \label{tab:table4b}
    \begin{tabular}{lcc}
        \toprule
        & \multicolumn{2}{c}{\textbf{Test}} \\
        \cmidrule(lr){2-3}
        \textbf{Train} & \textbf{ELRA-S0329} & \textbf{EmoMatchSpanishDB} \\
        \midrule
        ELRA-S0329        & 98.38\%  & 35.66\% \\
        EmoMatchSpanishDB & 41.54\%  & 64.20\% \\
        \bottomrule
    \end{tabular}
\end{table}

\begin{table}[!t]
    \centering
    \caption{Classification report training ELRA-S0329 and testing EmoMatchSpanishDB}
    \label{tab:table6}
    \begin{tabular}{lccc}
        \toprule
        \textbf{Label} & \textbf{Precision} & \textbf{Recall} & \textbf{F1-score} \\
        \midrule
        ANGER    & 1.000   & 0.010   & 0.010 \\
        DISGUST  & 0.350   & 0.050   & 0.080 \\
        FEAR     & 0.380   & 0.510   & 0.440 \\
        JOY      & 0.200   & 0.220   & 0.210 \\
        NEUTRAL  & 0.350   & 0.870   & 0.500 \\
        SADNESS  & 0.110   & 0.010   & 0.010 \\
        SURPRISE & 0.610   & 0.350   & 0.450 \\
        \bottomrule
    \end{tabular}
\end{table}

\begin{table}[!t]
    \centering
    \caption{Classification report training EmoMatchSpanishDB and testing ELRA-S0329}
    \label{tab:table7}
    \begin{tabular}{lccc}
        \toprule
        \textbf{Label} & \textbf{Precision} & \textbf{Recall} & \textbf{F1-score} \\
        \midrule
        ANGER    & 0.480   & 0.510   & 0.490 \\
        DISGUST  & 0.350   & 0.420   & 0.380 \\
        FEAR     & 0.420   & 0.540   & 0.470 \\
        JOY      & 1.000   & 0.000   & 0.010 \\
        NEUTRAL  & 0.610   & 0.150   & 0.240 \\
        SADNESS  & 0.330   & 0.680   & 0.440 \\
        SURPRISE & 0.550   & 0.620   & 0.580 \\
        \bottomrule
    \end{tabular}
\end{table}

\clearpage

\section{Conclusions}

For ELRA-S0329 the three models proposed outperform the current state-of-the-art models. If we compare DeepSpectrum-VGG16 MSF-SVC, we obtain an accuracy increase of 4,51\%. and DeepSpectrum-FN and DeepSpectrum-AM obtain an accuracy increase of 7,43\% and 8,33\% versus RNN (MFCC + MS). As the performance is so outstanding, the classification report gives us not any new information.

For EmoMatchSpanishDB, only two of the models outperform the current state-of-the-art. DeepSpectrum-FN and DeepSpectrum-AM obtain an accuracy increase of 1,7\% and 4,1\% versus SVC. In the confusion matrix and the classification report we can observe that for emotions like anger or fear the system has both a high precision and recall, while for emotions like joy the performance is not so good.

In the cross-datasets experiments, we see that the model trained in EmoMatchSpanishDB outperforms the model trained in ELRA-S0329 in a 5,88\%. This was expected as EmoMatchSpanishDB, having 50 speakers instead of the 2 in ELRA-S0329, is more appropriate for in-the-wild emotion recognition.

If we analyze the confusion matrix in both cases, we see the model trained in ELRA-S039 has the highest performance for neutral labelled samples, not surprising considering the five subcategories present in the dataset. Instead, in the model trained in EmoMatchSpanishDB, this happens with emotions like Surprise and Anger.

Finally, we can affirm that DeepSpectrum has proven its usefulness in the field of emotion recognition. It allows us to apply image recognition developments and to do transfer learning to an audio recognition problem, furthering the available ways of improvement in the research field. 

Future lines of research should be directed to, in one hand, increasing the number of available datasets to do research, taking into account that languages like Spanish have a large number of speakers but a not so large number of datasets, compared to English. On the other hand, Generative Adversarial Networks (GANs)~\cite{hortal2021} and the addition of other audio parameters like MSFs~\cite{wu_automatic_2011} to the system could increase the performance of it.

\section*{Acknowledgments}
 This work is part of the project PLEC2021-007868, funded by MCIN and by the EU, and the project PID2022-138721NB-I00, funded by MCIN.

\clearpage

\bibliographystyle{splncs04}
\bibliography{egbib}
\end{document}